\documentclass[pra, superscriptaddress]{revtex4}

\begin{document}

%\doi{10.1080/0950034YYxxxxxxxx}
%\issn{1362-3044}
%\issnp{0950-0340} \markboth{E. Andersson, J. D. Cresser and M. J. W. Hall}{Finding the Kraus decomposition from a master equation and vice versa}

\title{Finding the Kraus decomposition from a master equation and vice versa}
\author{ERIKA ANDERSSON$^*$, JAMES D. CRESSER$^\dagger$ and  MICHAEL J. W. HALL$^\ddag$
\thanks{$^\ast$Corresponding author. Email: erika@phys.strath.ac.uk}
\\
\vspace{6pt} 
$^*$Department of Physics, SUPA, University of Strathclyde, Glasgow G4 0NG, UK\\
$^\dagger$Centre for Quantum Computer Technology, Department of Physics, Macquarie University, Sydney NSW 2109, Australia\\
$^\ddag$Theoretical Physics, IAS, Australian National University, Canberra ACT 0200, Australia\\
%\date{\today}
%\vspace{6pt}\received{31 October 2006}
}

\begin{abstract}
For any master equation which is local in time, whether Markovian, non-Markovian, of Lindblad form or not, a general procedure is given for constructing the corresponding linear map from the initial state to the state at time $t$, including its Kraus-type representations.  Formally, this is equivalent to solving the master equation. For an $N$-dimensional Hilbert space it requires (i) solving a first order $N^2\times N^2$ matrix time evolution (to obtain the completely positive map), and (ii) diagonalising a related $N^2\times N^2$ matrix (to obtain a Kraus-type representation).  Conversely, for a given time-dependent linear map, a necessary and sufficient condition is given for the existence of a corresponding master equation, where the (not necessarily unique) form of this equation is explicitly determined. It is shown that a `best possible' master equation may always be defined, for approximating the evolution in the case that no exact master equation exists. Examples involving qubits are given.
\end{abstract}

\maketitle

%\end{document}

\section{Introduction}

The theory of open quantum systems is concerned with the study of quantum systems whose dynamics is determined both by interactions internal to the system, and by influences that lie outside the system.
These external influences are, of course, fundamentally quantum mechanical in nature, but under some circumstances it is useful to suppose that they take the form of classical, stochastically varying driving terms in the Hamiltonian of the system.  Either way, what is typically dealt with is the master equation, the equation of motion for $\rho$, the density operator of the system.

In the case in which the system of interest is interacting with an external quantum system (typically referred to as the environment), use of the Nakajima-Zwanzig projection operator technique \cite{Naka,20} shows that under fairly general conditions, the master equation for $\rho=Tr[\chi]$, where $\chi$ is the density operator for the combined system plus environment, takes the form 
\begin{equation}
	\dot{\rho}(t)=-\frac{\rm i}{\hbar}\left[H,\rho(t)\right]
	+\int_{0}^{t}\mathcal{K}_{t,s}[\rho(s)]\,ds,
	\label{genmaster}
\end{equation}
where $H$ is the system Hamiltonian in the absence of the
interaction with the environment, and $\mathcal{K}_{t,s}$, the memory
kernel, is a linear map describing the effects of the environment on
the system.  The first term represents the coherent evolution of the
system, while the second term gives rise to incoherent effects such as
damping and fluctuations of the system.  These effects of the
environment are seen to enter via an integral over the past history of
the state of the system, implying that the evolution of this state at
time $t$ depends on its previous history and as such, the system is
loosely said to possess non-Markovian dynamics.  However, in many
applications an important approximation, the Born-Markov
approximation, can be made to (\ref{genmaster}), which amounts to
approximating the memory kernel above by
$$
\mathcal{K}_{t,s}[\rho(s)]\approx\mathcal{K}\delta(t-s)\rho(s)
$$
In this case, there is no memory, and the system dynamics is said to be
Markovian.  

Of importance for the work reported in this paper is the fact that it
has been shown via the time-convolutionless projection operator method
\cite{Chat,Shib} that the general master equation (\ref{genmaster})
can be written in a local-in-time form,
\begin{equation}
	\dot{\rho}(t)=\Lambda_{t}[\rho(t)],
	\label{timelocal}
\end{equation}
where $\Lambda_{t}$ is, amongst other requirements, a linear map such
that $\Lambda_{t}(\rho)$ is Hermitean and traceless for all
$\rho$.  In the particular case of Markovian dynamics, $\Lambda$ is
independent of time (provided $H$ is assumed to be time
independent), and can be shown to take a specific structural form
known as the Lindblad form \cite{Lindblad}.  The Markovian case is
much studied; much less is known about the general mathematical
character of non-Markovian master equations, or their physical
interpretation.  Known examples of non-Markovian master equations include those describing the damped harmonic oscillator and spontaneous decay of a two-level system \cite{BreuerPetroBook}.
This is a research area that is becoming of increasing importance with the recognition that error correction in
quantum computing might not be adequately handled using Markovian methods, as the assumptions underlying the Markov approximation seem to be incompatible with the requirements assumed for error correction in quantum computing \cite{Lidar}. The errors that occur in quantum computing need not be statistically independent, i.e. the error-producing process may have memory, hence the need for an understanding of non-Markovicity. In \cite{terhal, aharonov, aliferis}, non-Markovian noise models for fault-tolerant quantum error correction are considered.
A description in terms of time-local master equations seems to be necessary for constructing a quantum trajectory unravelling, and perhaps a measurement interpretation, of non-Markovian master equations \cite{carmichael, plenio, kleinekat, intravaia, Breuer, breuer2, breuer3}.

An important feature of the derivation of the result (\ref{timelocal}) is that it relies on the existence of a certain operator inverse; e.g. see \cite{Breuer}.  It is possible, at certain critical times, for this inverse to cease to exist.  Some initial states evolve to the same final state at this time, and beyond this time there is no local-in-time master equation.  This seems to be a common feature of the master equation description of non-Markovian dynamics. 
The existence of an operator inverse is found to reoccur in somewhat different form in the work to be described in this paper. We are able to show that a time-local master equation may exist even though the time evolution is not invertible, provided that certain consistency conditions are satisfied. 

In the case in which there is an external classical stochastic
influence, the density operator is obtained by averaging over all the
realizations of the stochastic driving term.  While it is reasonable
to expect that the master equation so obtained will be of the general
form (\ref{genmaster}), no such proof appears to be available, though
specific examples have been studied \cite{Cresser, Budini1, Budini2}.
Once again this equation can be written in the time-local form
(\ref{timelocal}).

An alternative way of representing the dynamics of an open quantum system
is based on the recognition that the density operator $\rho(t)$ must
be related to its initial value $\rho(0)$ by a (usually completely
positive) linear map, $\rho(t)=\phi_{t}[\rho(0)]$ \cite{kraus}.  This observation
provides us with an alternate proof that memory-kernel master
equations can in fact be rewritten in time-local form.  In particular,
consider some evolution process described by the linear map $\rho(t) =
\phi_t[\rho(0)]$, which satisfies a memory-kernel master equation of
the general form above (assuming the Hamiltonian evolution is absorbed
into $\mathcal{K}$), i.e., 
\[ \dot{\rho}(t) = \int_0^t ds\,\mathcal{K}_{s,t}[\rho(s)]\] 
for some linear kernel map
$\mathcal{K}_{s,t}$.  If it is assumed that the map $\phi_t$ is {\it
invertible} for the time interval considered, i.e., that there exists a linear map $\phi^{-1}_t$
satisfying $\phi^{-1}_t\circ \phi_t=I$ where $I$ is the identity map
$I(X)=X$, then one can write
\begin{eqnarray*}
\dot{\rho}(t) &=& \int_0^t ds\, (\mathcal{K}_{s,t}\circ\phi_s)[\rho(0)]\\
&=& \int_0^t ds\,
(\mathcal{K}_{s,t}\circ\phi_s\circ\phi^{-1}_t)\{\phi_t[\rho(0)]\}\\
&=& \Lambda_t[\rho(t)],
\end{eqnarray*}
where $\Lambda_t:= \int_0^t ds\,
\mathcal{K}_{s,t}\circ\phi_s\circ\phi^{-1}_t$, which is of time-local
form.  Note that the assumption that the evolution is invertible is
violated only if two initially distinct states evolve to the same
density operator at some time $t$ (e.g., if an `equilibrium state' is reached
within a finite time).  This is the same kind of behaviour commented
on earlier with regard to equation (\ref{timelocal}).  As we will show, even in this
case it is sometimes possible to describe the evolution by a
time-local master equation. Let us also note that the restriction that the time evolution be invertible need not be serious. Equilibrium states are typically not reached in a finite time in physical systems. Physically relevant examples where the time evolution is invertible for any finite time include spontaneous emission and optical phase diffusion \cite{Jimphase}. The latter is also an example of a time-local master equation where the time evolution is not of standard Lindblad form. 

This now brings us to the issue we wish to investigate in this paper:
the relationship between these two ways of describing open
system dynamics, i.e., the relationship between $\Lambda_t$ and $\phi_t$.  
This investigation is undertaken in two directions.
First, if we are given a specific form for $\Lambda_t$, i.e.\ we are
given the master equation -- which need not be Markovian so
$\Lambda_{t}$ is not necessarily of Lindblad form -- what is the
corresponding linear map $\phi_{t}$?  The inverse question is also
analysed: given a linear map $\phi_{t}$, is there a corresponding
$\Lambda_{t}$, and what is its form?

We find that the first question can be answered by directly showing
that given an arbitrary master equation $\dot{\rho} = \Lambda_t(\rho)$
for some linear map $\Lambda_t$, we can explicitly construct Kraus
operators $\{ A_k\}$ \cite{kraus} such that
\begin{equation} \label{krauseqn}
\rho(t) = \phi_t[\rho(0)] = \sum_k \epsilon_k(t) A_k(t) \rho(0)
A^\dagger_k(t)
\end{equation}
for a linear map $\phi_t$, where $\epsilon_k(t)\in \{-1,1\}$ and where
the map $\phi_t$ is completely positive if and only if $\epsilon_k(t)=1$
for all $k$.  The construction proceeds by first selecting an
orthonormal basis of Hermitian operators, which maps the master
equation to a linear matrix equation.  This is a procedure that has
been used elsewhere in the analysis of properties of master equations
\cite{BreuerPetroBook}.  The matrix equation is then solved to obtain
$\phi_t$, and finally, a result of Choi \cite{Choi, caves} is used to
construct a Hermitian matrix which, when diagonalised (or even if only
expressed as some sum of vector outer products) allows one to obtain
the corresponding Kraus-type representation(s).  An example of the
method applied to a non-Markovian master equation is
presented.

The inverse question turns out to be more interesting.  A crucial step
in the analysis relies on the existence of the inverse of a certain
matrix.  If this inverse exists for all time, the construction of the
master equation is straightforward, though not necessarily simple.
However the inverse may not exist, either for all time, or at least at
some critical time, or times.  In such cases, it is nevertheless
possible, provided certain consistency conditions are satisfied, to
introduce a generalised matrix inverse, the so-called Moore-Penrose
inverse, which leads to a master equation, which, however, is not
necessarily unique.  Even if the consistency conditions are not satisfied, so that no `true' master equation exists, it is possible, using this procedure, to define a `best possible' master equation. Application of the method, including a discussion
of the possible circumstances under which the generalised approach
must be followed, is illustrated by a simple example.

\section{The master equation in matrix form}
\label{matrixeqnsec}

The first step in relating a general, non-Markovian master equation
$\dot{\rho}(t)=\Lambda[\rho(t)]$ to a corresponding completely positive map
$\rho(t)=\phi_{t}[\rho(0)]$ relies on expressing this relationship in
matrix form.  The resultant equation is then the starting point for
the two investigations outlined in the Introduction.

Let $\{G_a\}$ denote any convenient orthonormal basis set for the
Hermitian operators on the Hilbert space, i.e.
\[
G_a^\dagger = G_a,~~~~~{\rm tr}[G_aG_b] = \delta_{ab}.
\]
There is no need to restrict these operators to be Hermitian, but
doing so gives a real representation for density operators.  It
follows that any operator $X$ can be expressed as
\[
X = \sum_j x_j G_j={\bm x^{\rm T} \bm G}
,~~~~~x_j= {\rm tr}[G_jX],
\]
where the last equation follows via multiplication of the first by
$G_k$ and taking the trace.

In particular, for any map $\phi$ (though we will be concerned mainly
with completely positive maps below) one has, expressing both $\rho$
and $\phi(\rho)$ as a basis sum, 
\[
\phi(\rho) = \sum_k {\rm tr}\left[ 
\phi(\sum_l {\rm tr}[G_l\rho ]G_l)\right] G_k = \sum_{k,l}{\rm
tr}[G_k\phi(G_l)] {\rm tr}[G_l\rho ] G_k,
\]
which can be put in the matrix form
\begin{equation} \label{phifr}
\phi(\rho) = ({\bf F}{\bm r})^{\rm T} {\bm
G},
\end{equation}
where 
\begin{equation} \label{fdef}
F_{kl}:= {\rm
tr}[G_k\phi(G_l)], ~~~~r_l:= {\rm tr}[G_l\rho ] . 
\end{equation}
Indicating time-dependence of $\phi$ and ${\bf F}$ via $\phi_t$ and ${\bf F}(t)$, and defining
\[
\rho(t):= \phi_t[\rho(0)],
\]
it follows that the time evolution of $\rho$ can be expressed as
\[
\dot{\rho} = [\dot{{\bf F}}(t) {\bm r}(0)]^{\rm T} {\bm G}.
\]

Now, suppose that $\rho(t)$ satisfies a master equation of the form
\begin{equation} \label{master}
\dot{\rho} = \Lambda(\rho).
\end{equation}
Note that the linear map $\Lambda$ may be time-dependent.
To be a valid master equation the linear map $\Lambda$ will have to satisfy certain
conditions [e.g., $\Lambda(\rho)$ must be Hermitian and traceless for
all $\rho$], which will be assumed to be satisfied here (at present all one actually needs is
the linearity of $\Lambda$).  Defining the matrix ${\bf L}$ by
\begin{equation} \label{ldef}
L_{kl} := {\rm tr}[G_k\Lambda(G_l)]
\end{equation}
implies immediately that this master equation can be rewritten as
\[
\dot{\rho} = [{\bf L} {\bm r}(t)]^{\rm T} {\bm G}.
\]
Note that ${\bf L}$ (and ${\bf F}$) are necessarily real, but not symmetric in
general.

Comparing the results of the last two paragraphs gives
\[
\dot{{\bf F}} {\bm r}(0) = {\bf L} {\bm r}(t) = {\bf L} {\bf F} {\bm r}(0).
\]
Hence, since by linearity this equation must hold for all vectors
${\bm r}(0)$ (whether or not they correspond to a density operator), and
noting that the $G$-decomposition is unique, one has the general
relation
\begin{equation}
	\dot{{\bf F}} = {\bf L}{\bf F} . \label{MainEquation}
\end{equation}
This equation provides a matrix representation of the corresponding relation $\dot{\phi}=\Lambda\circ\phi$, between the linear maps $\phi$ and $\Lambda$, and allows us to proceed in either of two directions.

First, for a given master equation one can determine the
corresponding linear map $\phi_t$.  In particular, the matrix ${\bf L}$ is calculated via Eq.~(\ref{ldef}); the matrix equation 
(\ref{MainEquation}) is then solved for ${\bf F}$, subject to the initial condition ${\bf F}(0)\equiv
{\bf I}$; and $\phi_t$ follows via the representation
\begin{equation} \label{phirep}
\phi_t(X) = ({\bf F} {\bm x})^{\rm T} {\bf G}.
\end{equation}
In fact, solving the matrix equation formally is trivial: one has
\begin{equation} \label{fsolution}
{\bf F}(t) = {\bf F}(0)~ {\cal T}\exp[\int_0^t ds\,{\bf L}(s)] = {\cal T}\exp[\int_0^t
ds\,{\bf L}(s)]
\end{equation}
(${\cal T}$ denotes time ordering).  Note that for
time-independent master equations this reduces to
\[
{\bf F}(t) ={\rm e}^{{\bf L}t}
\]
which can be evaluated using the standard technique of reducing ${\bf L}$ to
Jordan canonical form.  Details of the procedure to be followed to
construct a Kraus-type decomposition of $\phi_{t}$ from knowledge of ${\bf L}$
and hence ${\bf F}$, as per Eq.~(\ref{krauseqn}), are presented in section \ref{krausgensec}.

Conversely, if we are instead given the map $\phi_{t}$, we can construct the 
matrices ${\bf F}$ and $\dot{{\bf F}}$ via Eq.~(\ref{fdef}), and attempt to solve (\ref{MainEquation}) for ${\bf L}$, so as to determine a corresponding master equation.  This is straightforward when ${\bf F}$ is invertible, but leads to non-trivial 
considerations otherwise, and will be discussed in section \ref{cptomastersec}.

\section{Generating Kraus-type decompositions}
\label{krausgensec}

Choi has shown that complete positivity of $\phi$ is equivalent to
the positivity of a particular matrix ${\bf S}$ \cite{Choi, caves}, and that the
Kraus decompositions of $\phi$ are related to the outer product
decompositions of ${\bf S}$.  The master equation gives us a matrix ${\bf L}$,
which in turn gives us a matrix ${\bf F}$, characterising the linear evolution map $\phi_t$.  
In order to proceed from ${\bf F}$ to a Kraus
decomposition, all we need to do is find ${\bf S}$ from ${\bf F}$, and then
diagonalise it.  This procedure works just as well for maps which are not completely positive, leading to a Kraus-type decomposition of the form of Eq.~(\ref{krauseqn}), as will be shown below. In fact, our construction of {\bf S} from {\bf L} allows us to {\it determine} whether or not, for a given proposed master equation, the corresponding map is completely positive.

The matrix ${\bf S}$ is defined as follows.  Let $\{|\alpha\rangle\}$ denote an
orthonormal basis for the Hilbert space in question, with $\alpha=1,2,\ldots,N$, and for a given
linear map $\phi$ define the $N^2\times N^2$ matrix
\begin{equation}
S_{ab} := \langle\alpha_1|\, \phi(|\alpha_2\rangle\langle \beta_2 |)\, |\beta_1\rangle .
\label{Sdef}
\end{equation}
Here the indexes $a$ and $b$ denote the ordered pairs $(\alpha_1,\alpha_2)$ and
$(\beta_1,\beta_2)$ respectively. We will refer to ${\bf S}$ as the Choi representation
of $\phi$ \cite{Choi}.

Noting that the $\tau_a\equiv%:=
|\alpha_1\rangle\langle \alpha_2|$ form an orthonormal (non-Hermitian) basis for the operators on
the Hilbert space, one can write (similarly as for the $G_a$
basis)
\begin{eqnarray}
	\phi(\rho) &=& \sum_{ab} S_{ab} \, \langle \alpha_2|\rho|\beta_2\rangle
	\, |\alpha_1\rangle\langle \beta_1| \nonumber \\
	&=& \sum_{ab} S_{ab} \, |\alpha_1\rangle\langle \alpha_2 |\, \rho\,
	|\beta_2\rangle \langle \beta_1| \nonumber \\
	&=& \sum_{ab} S_{ab} \tau_a \, \rho\, \tau_b^\dagger  \label{choirep} .
\end{eqnarray}
It follows trivially that if $\phi$ maps Hermitian operators to
Hermitian operators, then ${\bf S}$ is Hermitian, i.e., 
\[
S^*_{ba} = S_{ab}.
\]
Choi further demonstrated that {\it $\phi$ is completely positive if
and only if ${\bf S}$ is positive, i.e., ${\bf S}\geq 0$} \cite{Choi, caves}.  

The Hermitian property implies that one can always decompose ${\bf S}$ as a  sum of
outer products, i.e., 
\begin{equation} \label{sdecomp}
{\bf S} = \sum_i \epsilon_i {\bm V}(i) \,{\bm V}(i)^\dagger
\end{equation} 
for some
set of vectors $\{ {\bm V}(i)\}$.  For example, if the eigenvalues and
normalised eigenvectors of ${\bf S}$ are given by 
\begin{equation} \label{eig1}
{\bf S} {\bm X}(i) = \lambda_i {\bm X}(i),
\end{equation}
where the $\lambda_i$ are not necessarily positive as we are here not
assuming $\phi_t$ to be completely positive, we may then take
\begin{equation} \label{eig2}
{\bm V}(i) = \sqrt{|\lambda_i|} {\bm X}(i) ,~~~~\epsilon_i = {\rm sign}(\lambda_i) ,
\end{equation}
corresponding to the {\it orthogonal} decomposition 
\[
{\bf S} = \sum_i \lambda_i {\bm X}(i) \,{\bm X}(i)^\dagger 
\]
of ${\bf S}$.  Note that all
decompositions of ${\bf S}$ are related by unitary maps, and hence given
one, one can obtain all others \cite{Choi,caves}.

Each such decomposition of ${\bf S}$ corresponds to a Kraus-type decomposition of $\phi$
\cite{Choi, caves}. In particular, from Eqs.~(\ref{choirep}) and (\ref{sdecomp}) one has
\begin{equation}
\phi(\rho) = \sum_{i,a,b} \epsilon_i {\bm V}(i)_a \tau_a \,\rho\, \tau_b^\dagger 
 {\bm V}(i)_b^* = \sum_i \epsilon_i A_i\,\rho\,A_i^\dagger,
\end{equation}
as per Eq.~(\ref{krauseqn}), where the Kraus operators are given by
\begin{equation} \label{aidef}
A_i := \sum_a {\bm V}(i)_a\,\tau_a = {\bm V}(i)\cdot{\bm \tau}.
\end{equation}
This slightly generalises the usual Kraus representation for completely positive maps \cite{kraus}, for which case one has ${\bf S}\geq 0$ and hence that $\epsilon_i\equiv 1$ \cite{Choi, caves}.

The above result provides us with the tool for finishing off our problem.
Recall that the master equation gave us a matrix ${\bf L}$, which in turn
gave us a matrix ${\bf F}$, characterising the completely positive map
$\phi_t$ as per Eq.~(\ref{phirep}).  To finally obtain a Kraus-type decomposition, one only needs to determine ${\bf S}$ from ${\bf F}$, and then diagonalise it.

Determining ${\bf S}$ from ${\bf F}$ is quite straightforward, and relies on the link
between the two bases $\{G_a\}$ and $\{\tau_a\}$ for operators on the
Hilbert space.  Note first that 
\[
\phi_t(|\alpha_2\rangle\langle\beta_2|) 
=\sum_c {\rm tr}[|\alpha_2\rangle\langle\beta_2|G_c]\, \phi_t(G_c) =\sum_c
\langle\beta_2|G_c|\alpha_2\rangle\, \phi_t(G_c),
\]
and that
\[
\phi_t(G_c) =\sum_{rs} {\rm tr}[G_c G_r] F_{sr} G_s = \sum_s F_{sc} G_s.
\] 
Then, from the definition of ${\bf S}$ in equation (\ref{Sdef}), one has
\begin{eqnarray}
	S_{ab} &=& \langle\alpha_1| \phi_t(|\alpha_2\rangle\langle\beta_2|) |\beta_1\rangle \nonumber\\
	&=& \sum_{c,s} \langle\alpha_1|F_{sc}G_s|\beta_1\rangle\, 
	\langle \beta_2|G_c|\alpha_2\rangle\nonumber\\
	&=& \sum_{r,s} F_{sr} \langle\beta_2|G_r|\alpha_2\rangle\, 
	\langle \alpha_1|G_s|\beta_1\rangle\nonumber\\
	&=& \sum_{r,s} F_{sr}\, {\rm tr}[G_r \tau_a^\dagger G_s \tau_b] \label{sfromf}.
\end{eqnarray}
This gives the explicit connection between ${\bf F}$ and ${\bf S}$ as desired.

For calculational purposes it can be convenient to consider a unitary transformation
\begin{equation}
{\bf S}^{(W)}:= {\bf W}^\dagger {\bf SW} 
\end{equation}
of the Choi matrix, for some unitary matrix ${\bf W}$.  Note that this transformation defines an associated orthonormal basis set $\{ H_a\}$, for operators on the Hilbert space, via
\[   H_a := \sum_b W_{ab} \tau_b , \]
and it follows that the elements of ${\bf W}$ and ${\bf S}^{(W)}$ are given by [recalling $b\equiv(\beta_1,\beta_2)$]
\begin{equation} \label{wdef}
W_{ab} = {\rm tr}[H_a \tau^\dagger_b] = \langle \beta_1|H_a|\beta_2\rangle,~~~~~{\bf S}^{(W)}_{ab} = \sum_{rs} F_{sr}\, {\rm tr}[G_r H_a^\dagger G_s H_b] .
\end{equation}
The calculation of ${\bf S}$ may, on occasion, be simplified by choosing an appropriate basis $\{ H_a\}$, and using equation (\ref{wdef}) to first determine ${\bf W}$ and ${\bf S}^{(W)}$, to obtain the alternative decomposition
\begin{equation} \label{choireph}  
\phi(\rho) = \sum_{a,b} {\bf S}^{(W)}_{ab} H_a\rho H_b^\dagger  .
\end{equation}
Note that the Choi matrix ${\bf S}$ is positive if and only if ${\bf S}^{(W)}$ is positive.

\section{An example: unital qubit evolution}

It may be shown, noting that any orthogonal basis set for the traceless operators on a 2-d Hilbert space are related to the Pauli spin operators $\{ \sigma_j\}$ by some unitary transformation, that any map $\Lambda_t$ generating a time-local master equation for a qubit can be put in the canonical form
\begin{equation} 
 \Lambda_t(X) = \sum_{j=1}^3 \gamma_j \left(
\sigma^U_j X\sigma^U_j - X\right)  + \frac{1}{2} {\rm tr}[X]\sum_{j=1}^3 \alpha_j\sigma^U_j , 
\end{equation}
where $\sigma^U_j:={\bf U}(t)\sigma_j {\bf U}(t)$ for some unitary matrix ${\bf U}(t)$, and the $\gamma_j$ and $\alpha_j$ are functions of time.  Any Hamiltonian term has been absorbed by transforming to a suitable `interaction picture' (which acts to redefine ${\bf U}(t)$ appropriately and leaves $\gamma_j$ and $\alpha_j$ invariant). 

The evolution described by $\partial\rho/\partial t=\Lambda_t(\rho)$ is defined to be {\it unital} if the maximally mixed state $\frac{1}{2}\hat{1}$ is a fixed point, i.e., if $\Lambda_t(\hat{1})\equiv 0$.  Unital evolution is thus equivalent to the choice $\alpha_j(t)\equiv 0$ in the above equation.  Note that, for unital evolution, the master equation is Lindblad-type if and only if $\gamma_j(t)\geq 0$ for all $j$ (although this will not be assumed in what follows). With a Lindblad-type master equation, we mean a master equation that resembles Lindblad form, but has time-dependent decay rates \cite{BreuerPetroBook}.

As a simple application of the methods in sections 2 and 3 we will consider the particular case of unital evolution where ${\bf U}$ is time-independent. By a suitable choice of qubit basis states $\{|1\rangle,|2\rangle\}$, we can take ${\bf U}\equiv \hat{1}$ without any loss of generality, yielding the corresponding class of  master equations
\[ %\frac{\partial\rho}{\partial t} 
\dot{\rho}(t) = \gamma_1\sigma_1\rho\sigma_1 + \gamma_2\sigma_2\rho\sigma_2 + \gamma_3\sigma_3\rho\sigma_3 - (\gamma_1 + \gamma_2 +\gamma_2)\rho . \]
This class includes, for example, the physically relevant example of a two-level atom driven by a phase noisy laser, when no decorrelation assumptions are made \cite{Jimphase}.  In this case, the master equation is of the form
\[
\dot{\rho}(t) = \frac{\dot{g_1}}{2g_1}(\rho-\hat{\sigma}_+\rho\hat{\sigma}_--\hat{\sigma}_-\rho\hat{\sigma}_+)+
\left(\frac{\dot{g}_2}{g_2}-\frac{\dot{g}_1}{2g_1}\right)
(\hat{\sigma}_+\hat{\sigma}_-\rho\hat{\sigma}_-\hat{\sigma}_+
+\hat{\sigma}_-\hat{\sigma}_+\rho\hat{\sigma}_+\hat{\sigma}).
\]
Here $\sigma_+=|2\rangle\langle 1|$ and  $\sigma_-=|1\rangle\langle 2|$, where $|1\rangle$ and $|2\rangle$ are the lower and upper states of the atom, and the functions $g_1$ and $g_2$ are defined in terms of their Laplace transforms
\[\tilde{g}_1 = \frac{s+D}{s^2+Ds+4\lambda^2}, 
~~~\tilde{g}_2 = \frac{(s+D)(s+4D)+2\lambda^2)}{s[(s+D)(s+4D)+2\lambda^2] + 2\lambda^2(s+4D)},
\]
where $D$ is a diffusion rate and $\lambda$ is the strength of the coupling between the two levels due to the phase-fluctuating classical field \cite{Jimphase}.

Choosing the basis set $\{G_a\}$ via $G_a =\sigma_a/\sqrt{2}$, with $\sigma_0:= \hat{1}$, the matrix ${\bf L}$ corresponding to the master equation follows via equation (\ref{ldef}) as
\[ {\bf L}(t) = 
	 \left(\begin{array}{cccc}0 & 0 & 0 & 0 \\
	0 & -\gamma_2(t) -\gamma_3(t) & 0 & 0 \\
	0 & 0 & -\gamma_1(t) -\gamma_3(t) & 0 \\
	0 & 0 & 0 & -\gamma_1(t) -\gamma_2(t)\end{array}\right)  . \]
This matrix is diagonal (for non-unital maps one has $L_{j0}=\alpha_j$), and hence equation (\ref{fsolution}) immediately yields
\[ {\bf F}(t) = 
	 \left(\begin{array}{cccc}1 & 0 & 0 & 0 \\
	0 & \Gamma_1(t) & 0 & 0 \\
	0 & 0 & \Gamma_2(t) & 0 \\
	0 & 0 & 0 & \Gamma_3(t)\end{array}\right)   \]
for the evolution matrix ${\bf F}$, with
\[ \Gamma_i(t) := {\rm e}^{-\int_0^t ds\,\gamma_j(s)+\gamma_k(s)} , \]
where $\{i,j,k\}$ is a permutation of $\{1,2,3\}$. Note that $\Gamma_j(0)=1$ and $\Gamma_j(t)\geq 0$.

To find a corresponding Kraus-type decomposition, it is convenient to exploit properties of the Pauli spin matrices by choosing $H_a\equiv G_a$ in equation (\ref{wdef}), so that 
\[ {\bf S}^{(W)}_{ab} = \frac{1}{4}\sum_{r,s} {\bf F}_{sr}\,{\rm tr}[\sigma_r\sigma_a\sigma_s\sigma_b] .\]
One obtains
\[	{\bf S}^{(W)} = \frac{1}{2} {\rm diag}[1+\Gamma_1+\Gamma_2+\Gamma_3, 1+\Gamma_1-\Gamma_2-\Gamma_3,1-\Gamma_1+\Gamma_2-\Gamma_3,1-\Gamma_1-\Gamma_2+\Gamma_3],   \]
and hence, from equation (\ref{choireph}), that
\[ \phi_t(\rho) = \frac{1}{4} \left( 1+\Gamma_1+\Gamma_2+\Gamma_3\right)\rho + \frac{1}{4} \sum_{k=1}^3 \left[2\Gamma_j +1 -( \Gamma_1+\Gamma_2+\Gamma_3)\right] \, \sigma_j \rho \sigma_j .\]

It follows that the evolution is completely positive if and only if the diagonal elements of ${\bf S}^{(W)}$ are positive.  This is trivial for the first element, and hence complete positivity reduces to the conditions
 \begin{equation} \label{cpcond}
     \Gamma_i + \Gamma_j \leq 1 + \Gamma_k ,
 \end{equation}
where $\{i,j,k\}$ run over the cyclic permutations of $\{1,2,3\}$.  These inequalities, together with $\Gamma_j\geq 0$, determine a convex polyhedron in three dimensions, with vertices $(0,0,0)$, $(1,0,0)$, $(0,1,0)$, $(0,0,1)$ and $(1,1,1)$.  The conditions are satisfied automatically if $\gamma_j(t)\geq 0$ for all times, i.e. if the master equation is of Lindblad-type, but can also be satisfied in many other cases.  

Finally,  assuming that conditions (\ref{cpcond}) are satisfied, the density operator at time $t$ has the explicit Kraus representation
\[ \rho(t) = \sum_{i=0}^3 A_i\rho(0) A^\dagger_i ,\]
with 
\begin{eqnarray*}
A_0 &=& \frac{1}{2} (1+\Gamma_1+\Gamma_2+\Gamma_3)^{1/2}\hat{1}\\
A_1 &=& \frac{1}{2} (1+\Gamma_1-\Gamma_2-\Gamma_3)^{1/2}\sigma_1 \\
A_2 &=& \frac{1}{2} (1-\Gamma_1+\Gamma_2-\Gamma_3)^{1/2}\sigma_2\\
A_3 &=& \frac{1}{2} (1-\Gamma_1-\Gamma_2+\Gamma_3)^{1/2}\sigma_3 .
\end{eqnarray*}

\section{From linear maps to master equations} 
\label{cptomastersec}

We now wish to study the `inverse' question mentioned in the Introduction, that is, given a linear evolution map $\phi_{t}$, can we construct a corresponding master equation?  This is clearly equivalent to determining whether the matrix equation $\dot{{\bf F}} = {\bf L}{\bf F}$ in Eq.~(\ref{MainEquation}) has a suitable solution for ${\bf L}$, for a specified matrix function ${\bf F}(t)$.  In the case that
${\bf F}(t)$ is invertible one has the unique
solution 
\begin{equation} \label{invertible}
{\bf L}=\dot{{\bf F}}{\bf F}^{-1} ,
\end{equation}
and hence a unique corresponding master
equation.  

However, in the singular case, in which ${\bf F}^{-1}$ does not exist, the situation is not so clear.  For example, $\det {\bf F}=0$ in such a case, and hence from Eq.~(\ref{MainEquation}) one 
requires that $\det \dot{{\bf F}}=0$.  If the latter condition is not satisfied there
can be no solutions for ${\bf L}$, and hence no corresponding master equation.
On the other hand, as we will show below, there can be 
situations for which ${\bf F}$ is not invertible for which there are a {\it plurality} of master equations
which generate the same evolution. We will begin by discussing the general consistency conditions which must be satisfied for a time-local master equation to exist for a given evolution map $\phi_t$, and show how to obtain the corresponding master equation(s).  It will also be shown that, even when the consistency condition is not met, one can still define a  `best possible' master equation corresponding to the map.

\subsection{Consistency conditions}
\label{consistencysec}

A time-local master equation is first-order in time.  Hence, if the state is known at time $t$, the state at all later times is uniquely determined.  It follows in particular that if the trajectories of two density operators $\rho(t)$ and $\tau(t)$ intersect at some time $t=t'$, i.e., $\rho(t')=\tau(t')$, then the trajectories merge into a single trajectory for all later times, i.e., $\rho(t)=\tau(t)$ for $t>t'$.  This general causal requirement, that when two trajectories `kiss' they then stay together forever, might be termed a `fidelity' constraint on time-local master equations.  Note that since any merging of trajectories is irreversible, the evolution map $\rho(t)=\phi_t[\rho(0)]$ is not invertible in such a case. By explicit quantum trajectory simulations, Di\'{o}si et al. have numerically demonstrated this kind of behaviour \cite{diosi}.

It follows that for a given map $\phi_t$ to have a corresponding time-local master equation, the condition
   \[ \phi_{t'}[\rho(0)] = \phi_{t'}[\tau(0)] ~~{\rm implies}~~ \phi_{t}[\rho(0)] = \phi_{t}[\tau(0)]~~{\rm for~all}~~t\geq t' \]
must be satisfied for all density operators $\rho_0$ and $\tau_0$.  Noting equation (\ref{phifr}), this may be re-expressed in terms of the matrix ${\bf F}$ as
\[ {\bf F}(t')\left[ {\bm x}(0)-{\bm y}(0) \right] =0 ~~{\rm implies}~~  {\bf F}(t)\left[ {\bm x}(0)-{\bm y}(0) \right] =0    ~~{\rm for~all}~~t\geq t', \]
where ${\bm x}(0)$ and ${\bm y}(0)$ denote the representations of $\rho(0)$ and $\tau(0)$ with respect to the basis set $\{G_a\}$.  Hence, by linearity, any vector in the kernel of ${\bf F}$, ${\rm ker~}{\bf F}$, at some time $t'$ must remain in the kernel for all later times. Denoting the projection matrix onto the kernel of ${\bf F}(t)$ by ${\bf K}(t)$ (thus, ${\bf K}^2={\bf K}$ and ${\bf F}{\bf K}=0$), it further follows from the defining equation $\dot{{\bf F}}={\bf L}{\bf F}$ that $\dot{{\bf F}}{\bf K}=0$.  Hence, the fidelity constraint reduces to the two consistency conditions
\begin{equation} \label{FKcondition}
{\rm ker~}{\bf F}(t') \subseteq {\rm ker~}{\bf F}(t),~~~~~\dot{{\bf F}}(t)\,{\bf K}(t) =0
\end{equation}
on ${\bf F}(t)$, for all times $t\geq t'\geq 0$.  Note the first of these conditions implies that the dimension of the kernel of ${\bf F}$ increases monotonically, and hence provides a measure of irreversibility.  Note also that if ${\bf F}(t)$ is invertible for all times $t$, then ${\bf K}\equiv 0$, and the conditions are trivially satisfied.  In this case, the corresponding master equation is uniquely defined by the matrix ${\bf L}$ in equation (\ref{invertible}).

Finally, note that a third consistency condition has been implicitly assumed from the outset: that $\dot{\rho}$ is well defined at all times - otherwise, of course, no master equation of any kind can be defined.  This is equivalent to the assumption that the matrix $\dot{{\bf F}}(t)$ is well defined for all $t\geq 0$.
  
\subsection{Finding the master equation when the inverse of ${\bf F}$ does not exist}
\label{Lsolutionsec}

Even if the matrix ${\bf F}$ is not invertible, so that equation (\ref{invertible}) cannot be applied, a corresponding master equation can be shown to exist, provided that the consistency conditions (\ref{FKcondition}) hold.

In particular, let 
\[
{\bf F} = {\bf U}{\bf D}{\bf V}^{\rm T}
\]
denote the singular value decomposition of the real matrix ${\bf F}$.  Thus,
${\bf U}$ and ${\bf V}$ are orthogonal matrices which diagonalise ${\bf F}{\bf F}^\dagger$ and ${\bf F}^\dagger {\bf F}$ respectively, and ${\bf D}$ is a real diagonal matrix of the form 
\[
{\bf D}=diag[s_1, s_2,\dots,s_r, 0,\dots,0],
\]
where $r$ is the rank of ${\bf F}$ and $s_1\geq s_2\geq \dots \geq s_r>0$.
The singular values $s_1,s_2, \dots$ are the square roots of the non-zero eigenvalues of either of ${\bf F}{\bf F}^\dagger$ and ${\bf F}^\dagger {\bf F}$.  Further, define the related diagonal matrix
\[
\tilde{{\bf D}}:=diag[1/s_1, \dots,1/s_r, 0,\dots,0],
\]
and the projection matrix
\[
{\bf P} := \tilde{{\bf D}}{\bf D} = {\bf D}\tilde{{\bf D}} = diag[1,\dots,1,0,\dots 0] ={\bf P}^2 
\]
having $r$ 1s and $N^2-r$ 0s.  The matrix 
\begin{equation} \label{moorepenrose}
\tilde{{\bf F}}:= {\bf V}\tilde{{\bf D}}{\bf U}^{\rm T} 
\end{equation}
is called the Moore-Penrose inverse of ${\bf F}$ \cite{hornjohnson}.

If the consistency conditions (\ref{FKcondition}) are satisfied, a solution to the equation in Eq. (\ref{MainEquation}), $\dot{{\bf F}}={\bf L}{\bf F}$, is then given by
\begin{equation}
	{\bf L} = \dot{{\bf F}} {\bf V}\tilde{{\bf D}}{\bf U}^{\rm T}= \dot{{\bf F}}\tilde{{\bf F}}
	\label{Lgensolution}
\end{equation}
where $\tilde{{\bf D}}$ and $\tilde{{\bf F}}$ are defined above. Note that this reduces to equation (\ref{invertible}) when ${\bf F}$ is invertible, since $\tilde{{\bf F}}={\bf F}^{-1}$ in this case.

To show there can indeed be an {\it infinity} of solutions for ${\bf L}$ (and hence an infinity of master equations corresponding to the evolution map $\phi_t$), note first that
${\bf V}{\bf P}{\bf V}^{\rm T}$ is the projection onto the range of ${\bf F}$, and hence
\begin{equation} \label{kfromsvd}
{\bf K} = {\bf I} -{\bf V}{\bf P}{\bf V}^{\rm T}  .
\end{equation}
Now consider the ansatz
\begin{equation} \label{ansatz}
{\bf L} = \dot{{\bf F}}\tilde{{\bf F}} + {\bf M},
\end{equation}
for some matrix ${\bf M}$.  It follows immediately that
\[ {\bf L}{\bf F} = \dot{{\bf F}}\tilde{{\bf F}}{\bf F} + {\bf M}{\bf F} = \dot{{\bf F}}{\bf V}\tilde{{\bf D}}{\bf U}^{\rm T}{\bf U}{\bf D}{\bf V}^{\rm T} +{\bf M}{\bf F} = \dot{{\bf F}}{\bf V}{\bf P}{\bf V}^{\rm T} +{\bf M}{\bf F} = \dot{{\bf F}}({\bf I}-{\bf K})+{\bf M}{\bf F} = \dot{{\bf F}}+{\bf M}{\bf F} ,\]
where the last equality follows via the second consistency condition in equation (\ref{FKcondition}).  Hence, equation (\ref{MainEquation}) is satisfied if and only if 
\[ {\bf M}{\bf F} = 0. \]
The choice ${\bf M}=0$ gives the particular solution in equation
(\ref{Lgensolution}).  Note that one is forced to make this choice whenever ${\bf F}$ is invertible.

Finally, the explicit form of the master
equation $\dot{\rho}=\Lambda_t(\rho)$, corresponding to equation (\ref{Lgensolution}), is prescribed by
\begin{equation}
	\label{generalmastereqn}
	\Lambda_t(X) := \sum_{ab} R_{ab}(t) \tau_a\,X\,\tau_b^\dagger ,
\end{equation}
 where
\begin{equation} \label{rdef}
R_{ab}:= \sum_{rs} (\dot{{\bf F}}\tilde{{\bf F}})_{rs}\,{\rm
tr}[G_r\tau_a^\dagger G_s \tau_b],
\end{equation}
and where the $\tau_a$ were defined in section \ref{krausgensec}.
To obtain these expressions, note that $R_{ab}$ is just the Choi matrix
of the linear map corresponding to the solution for ${\bf L}$, analogous to equations (\ref{Sdef}) and (\ref{sfromf}).  
Note that one can in fact obtain a similar master equation for each allowable matrix ${\bf M}$ in equation (\ref{ansatz}), where each such master equation generates precisely the same evolution map $\phi_t$ (providing, of course, that all consistency conditions are satisfied).

\subsection{Best possible master equation when the consistency conditions are not met}

It is of interest to note that the Moore-Penrose inverse of ${\bf F}$ in equation (\ref{moorepenrose}) is always well-defined, whether or not the consistency conditions in section 5.1 are met.  Hence, the corresponding matrix ${\bf L}$ in equation (\ref{Lgensolution}), and the master equation map $\Lambda_t$ in (\ref{generalmastereqn}) are similarly always well-defined.  It turns out that, for {\it any} given evolution map $\phi_t$, the map $\Lambda_t$ defined by (\ref{generalmastereqn}) always yields the {\it best possible} time-local master equation (and yields an exact master equation in the particular case that the consistency conditions are satisfied). 

In particular, consider the problem of attempting to model a given evolution process, $\rho(t):=\phi_t[\rho(0)]$, by a time-local master equation.  For example, $\phi_t$ might be defined via a particular interaction of the system with a `bath', or obtained from phenomenological or theoretical considerations.  This problem may be formulated more precisely as determining the map $\Lambda_t$ for which
\begin{equation} \label{minimumnorm}
\| \dot{\phi_t} - \Lambda_t\| = {\rm ~minimum}, 
\end{equation}
where the norm of a given linear map $\phi$ is identified with the Hilbert-Schmidt norm of the corresponding matrix representation of $\phi$ in equation (\ref{fdef}), i.e.,
\[ \|\phi\| \equiv \| {\bf F}\| := \left({\rm tr}[{\bf F}^{\rm T} {\bf F}]\right)^{1/2} \]
(note that the norm is independent of the choice of basis set $\{G_a\}$).  
It follows that the problem is equivalent to determining the matrix ${\bf L}$ such that
\[ \|\dot{{\bf F}}-{\bf L}{\bf F}\| = {\rm ~minimum} . \]

It will be shown that the solution to the above problem, further satisfying the `least squares' property
\begin{equation} \label{squares}
 \| \Lambda_t\| = \| {\bf L}\| = {\rm minimum}, 
 \end{equation}
is given by ${\bf L}=\dot{{\bf F}}\tilde{{\bf F}}$, corresponding to the map in equation (\ref{generalmastereqn}).  This map is therefore, in this sense, the best possible for modelling the process $\phi_t$, even in the case where no true time-local master equation exists.  

In particular, noting that ${\bf K}({\bf I}-{\bf K})=0$ and ${\bf FK}=0$, one has
\begin{eqnarray*}
\|\dot{{\bf F}}-{\bf L}{\bf F}\|^2 &=& \| (\dot{{\bf F}}-{\bf L}{\bf F}){\bf K} + (\dot{{\bf F}}-{\bf L}{\bf F})({\bf{I}-\bf K})\|^2 \\
&=& \| (\dot{{\bf F}}-{\bf L}{\bf F}){\bf K}\|^2 + \|(\dot{{\bf F}}-{\bf L}{\bf F})({\bf{I}-\bf K})\|^2\\
&=& \| \dot{{\bf F}}{\bf K}\|^2 + \| \dot{{\bf F}}({\bf{I}-\bf K}) - {\bf L}{\bf F}\|^2 \\
&\geq& \| \dot{{\bf F}}{\bf K}\|^2 .
\end{eqnarray*}
Further, from equations (\ref{moorepenrose}) and (\ref{kfromsvd}) one has $\tilde{{\bf F}}{\bf F}= {\bf V}{\bf P}{\bf V}^{\rm T} = {\bf I}-{\bf K}$.  It follows immediately that the minimum value is obtained for the choice ${\bf L}=\dot{{\bf F}}\tilde{{\bf F}} +{\bf M}$, where ${\bf M}$ is any matrix which satisfies ${\bf MF}=0$, precisely as per equation (\ref{ansatz}). Note from ${\bf UPU}^{\rm T}{\bf F}={\bf F}$ that ${\bf M}={\bf M}({\bf I}-{\bf UPU}^{\rm T})$.  Thus, noting further that $\tilde{{\bf F}}{\bf UPU}^{\rm T} = \tilde{{\bf F}}$, one has
\[ \| {\bf L}\|^2 = \| \dot{{\bf F}}\tilde{{\bf F}}\|^2 + \|{\bf M}\|^2 . \]
Hence, the `least squares' solution is given by ${\bf L}=\dot{{\bf F}}\tilde{{\bf F}}$ as claimed.

Note that for qubits undergoing purely Hamiltonian evolution, with ${\rm i}\hbar \Lambda(\rho) = [H,\rho]$, one finds
\begin{equation} 
\hbar^2\|\Lambda\|^2 = {\rm tr}[H^2] . \end{equation}
This suggests it may be possible to interpret the `least squares' property (\ref{squares}) more generally as some kind of `minimum energy' property.

Finally, note from section 5.2 that the norm $\| \dot{\phi_t} - \Lambda_t\|$ vanishes identically for all times if the second consistency condition in equation (\ref{FKcondition}) is satisfied.  However, this does {\it not} imply that the {\it first} consistency condition is necessarily satisfed at all times.  In particular, the corresponding `best possible' master equation can only track the evolution map $\phi_t$ from $t=0$ only up until times for which the first consistency condition is met: no first-order equation in time can model the splitting of a given trajectory into two or more trajectories.

\section{From Kraus representation to master equation: minimal decoherence of a two-level atom}

Here the methods of the previous section are demonstrated for the case of a two-level atom, the evolution of which is described by a completely positive map of a particular form.  For this example the matrix ${\bf F}$ is not always invertible.  When it is not invertible, the consistency conditions for a master equation to exist may or may not be satisfied (and the master equation need not be unique). The master equation for a damped two-level atom given by Garraway \cite{barry} corresponds to a special case of the completely positive map we consider. We also note that this map can be obtained via a Jaynes-Cummings interaction with a single-mode field  \cite{wonderenlendi}. 

Using the notation of section 4, consider then a two-level atom (or qubit) with lower and upper levels $|1\rangle$ and $|2\rangle$ respectively.  It will be assumed that the probability of the atom being found in the upper level at time $t$, if initially in the upper level at time $t=0$, is given by some function $p(t)$.  Thus, $p(t)$ is the survival probability of the excited state, and lies between $0$ and $1$, with $p(0)=1$.  No other properties of the survival probability will be assumed for now, other than it is differentiable. Note that exponential decay into the lower level corresponds to the choice $p(t)= {\rm e}^{-\gamma t}$.  More generally, however, $p(t)$ may be non-monotonic, or even periodic, allowing for the possiblity of revivals.

We will assume that $\rho(t)$ is related to $\rho(0)$ by some linear map $\phi_t$. If we further allow for arbitrary decoherence of the off-diagonal elements of $\rho$, described by some function $f(t)$, then linearity and the conservation of probability imply the form
\begin{eqnarray}
\label{generalqubitevolution}
\phi_t(\rho) &=& p(t) \rho_{22} |2\rangle\langle 2| + \{\rho_{11} 
+[1- p(t)]\rho_{22}\}|1\rangle\langle 1|\nonumber\\
&~& + f(t)\rho_{21}|2\rangle\langle 1| + f^*(t) \rho_{12}|1\rangle\langle 2|
\end{eqnarray}
for all states $\rho$, where $f(t)$ is some complex function satisfying $f(0)=1$.  
The requirement that $\phi_t$ maps positive operators to positive operators implies that $\det \phi_t(\rho)\geq 0$ for all $\rho$, yielding
\[ 0\leq |f|^2\det\rho +\rho_{22}p(1-p\rho_{22}) - |f|^2\rho_{22}(1-\rho_{22}) \]
for all states $\rho$.  Taking $\rho=|1\rangle\langle 1|$ implies 
$|f|^2\leq p$, and it is easily checked that this condition is also sufficient to ensure positivity.  For simplicity, we will assume that strict equality holds, i.e., that $f(t)$ satisfies
\begin{equation} \label{pf2}
p(t) = |f(t)|^2 .
\end{equation}
This choice corresponds to maximising the off-diagonal elements of $\phi_t(\rho)$, i.e., to {\it minimal decoherence} between the levels for a given survival probability function.  

It may be checked that the above `minimal decoherence' map $\phi_t$ has the Kraus decomposition
$\rho(t) = \phi_t(\rho) = \sum_k A_k(t)\rho(0)A_k^\dagger (t)$ with
\begin{eqnarray}
\label{krausop}
A_1 (t)&=& |1\rangle\langle1| + f(t)|2\rangle\langle 2|\nonumber\\
A_2(t) &=& \sqrt{1-|f(t)|^2}|1\rangle\langle 2|. \label{Krausrep}
\end{eqnarray}
Hence the evolution is described by a completely positive map.  However, it will be seen that in general the corresponding master equation need not be of Lindblad-type form.

\subsection{Master equation when the inverses of $\phi$ and ${\bf F}$ exist}

To obtain the master equation, the matrix ${\bf F}$ first must be determined, and checked as to whether it is invertible  (and if it is not, whether the consistency conditions in section \ref{consistencysec} are satisfied). 
For simplicity, we will choose $f$ to be real. It can be shown that a non-zero imaginary part of $f$ would only add a term of Hamiltonian type to the resulting master equation. The basis $\{G_a\}$ is again chosen as
$G_{a}=\frac{1}{\sqrt{2}}\sigma_{a}$, as per section 4.  One finds
\begin{eqnarray}
\phi(G_0) &=& \sqrt{2}|1\rangle\langle 1| + f^2(t)G_3,~~
\phi(G_1) = f(t)G_1, ~~\nonumber\\
\phi(G_2) &=& f(t)G_2 ~~{\rm and}~~
\phi(G_3) = f^2(t) G_3.\nonumber
\end{eqnarray}
The matrix ${\bf F}$ and its time
derivative $\dot{{\bf F}}$ follow via Eq.~(\ref{fdef}) as
\begin{eqnarray}
{\bf F}= \left(\begin{array}{cccc}1 & 0 & 0 &0 \\
0 & f(t) & 0 & 0 \\
0 & 0 & f(t) & 0 \\
f^2(t)-1& 0 & 0 & f^2(t)\end{array}\right),~~~\nonumber\\
\label{fdotmatrix}
\dot{{\bf F}}= \left(\begin{array}{cccc}0 & 0 & 0 & 0 \\
0 & \dot{f}(t) & 0 & 0 \\
0 & 0 & \dot{f}(t) & 0 \\
2\dot{f}(t)f(t)& 0 & 0 & 2\dot{f}(t)f(t)\end{array}\right).
\end{eqnarray}

When $f(t)\neq 0$, then ${\bf F}$ is invertible, with
\begin{equation}
{\bf F}^{-1}= \left(\begin{array}{cccc}1 & 0 & 0 &0 \\
0 & 1/f(t) & 0 & 0 \\
0 & 0 & 1/f(t) & 0 \\
1/f^2(t)-1 & 0 & 0 & 1/f^2(t)\end{array}\right).
\end{equation}
The solution for ${\bf L}$ follows via Eq.~(\ref{invertible}) as
\begin{equation}
{\bf L} = \dot{{\bf F}}{\bf F}^{-1}
= \left(\begin{array}{cccc}0 & 0 & 0 & 0  \\
0 & \dot{f}(t)/f(t) & 0 & 0 \\
0 & 0 & \dot{f}(t)/f(t) & 0 \\
2\dot{f}(t)/f(t)& 0 & 0 & 2\dot{f}(t)/f(t)\end{array}\right).
\label{Lmatr}
\end{equation}

Finally, to obtain the master equation in the form $\dot{\rho}=\Lambda_t(\rho)$, we calculate the Choi matrix $\bf R$ in equation ({\ref{rdef}), corresponding to $\bf L$, yielding 
\begin{equation}
{\bf R}= \left(\begin{array}{cccc}2\dot{f}(t)/f(t) & \dot{f}(t)/f(t) & 0 &0 \\
\dot{f}(t)/f(t) & 0 & 0 & 0 \\
0 & 0 & 0 & 0 \\
0 & 0 & 0 & -2\dot{f}(t)/f(t)\end{array}\right),
\end{equation}
where the ordering $\{|2\rangle\langle 2|,|1\rangle\langle 1|,|2\rangle\langle
1|,|1\rangle\langle 2|\}$ for the basis elements has been chosen. The master equation immediately follows via equation (\ref{generalmastereqn}) as
\begin{eqnarray}
\label{2levelmastereqn}
	\dot{\rho}(t)=\Lambda_t[\rho(t)] 
	&=&\frac{\dot{f}(t)}{f(t)}\left(2|2\rangle\langle 2|\rho(t)|2\rangle\langle 2| 
	+|2\rangle\langle 2|\rho(t)|1\rangle\langle 1| 
	+|1\rangle\langle 1|\rho(t)|2\rangle\langle 2| 
	-2|1\rangle\langle 2|\rho(t)|2\rangle\langle 1| \right)\nonumber\\
	&=&-{\dot{f}(t)\over f(t)}\left(2\sigma_-\rho(t)\sigma_+ -\sigma_+
\sigma_-\rho(t)-\rho(t)\sigma_+ \sigma_- \right).
\end{eqnarray}
It may be shown that if we do not assume that $f$ is real, a term of Hamiltonian type is added to the master equation, so that it reads
\begin{eqnarray}
\dot{\rho}(t)
&=&-\frac{\dot{f}(t)f^*(t)+f(t)\dot{f}^*(t)}{2|f(t)|^2}\left(2\sigma_-\rho(t)\sigma_+ -\sigma_+
\sigma_-\rho(t)-\rho(t)\sigma_+ \sigma_- \right)
+\nonumber\\
&&\frac{\dot{f}(t)f^*(t)-f(t)\dot{f}^*(t)}{2|f(t)|^2}[\sigma_+\sigma_-,\rho(t)].
\label{mastereq}
\end{eqnarray}

For the case of exponential decay, i.e., $p(t)={\rm e}^{-\gamma t}$, we have $\dot{f}(t)/f(t)=-\gamma/2$, and
then the master equation is of standard Lindblad form, corresponding to Markovian behaviour \cite{BreuerPetroBook}. If $f(t)$ is time-dependent, but in a way so that  $\dot{f}(t)/f(t)$ is negative, then this corresponds to non-Markovian but still `Lindblad-type' behaviour. Such a master equation can unravelled along the same lines as when using the usual quantum trajectory methods \cite{zollerjumps, dum, montecarlo1, montecarlo2, carmichael}, as is done for instance in \cite{intravaia}.
If  $\dot{f}(t)/f(t)$ becomes positive, this corresponds to reabsorption and recoherence, and can be termed `anti-Lindblad' behaviour. For this case, other quantum trajectory unravelling techniques must be used \cite{BreuerPetroBook, kleinekat, Breuer}. 
If we have $|f(t_1)| < |f(t_2)|$, then the time evolution map from $\rho(t_1)$ to $\rho(t_2)$ is not completely positive.  The {\it total} time evolution from $t=0$ to $t=t_2$ is always completely positive, as indicated by the Kraus representation in (\ref{Krausrep}).  We will return to the issue of the relation between non-Lindblad master equations and complete positivity elsewhere.

\subsection{Master equation when the inverses of $\phi$ and ${\bf F}$ do not exist}

If at some time $t_0$ we have $f(t_0)=0$, i.e., the atom occupies the lower level with probability unity, then the inverse of the linear map $\phi$ does not exist at this time. This corresponds to the inverse of the matrix ${\bf F}$ not
existing. At such times ${\bf F}$ has the form
\begin{equation} \label{Ft0}
{\bf F} = \left(\begin{array}{cccc}1 & 0 & 0 & 0 \\
0 & 0 & 0 & 0 \\
0 & 0 & 0 & 0 \\
-1 & 0 & 0 & 0\end{array}\right),
\end{equation}
implying that ${\bf F}x=0$ for any vector with $x_0=0$.  The projection matrix ${\bf K}$ onto the kernel of ${\bf F}$ follows as
\begin{equation}
{\bf K}(t) = \left(\begin{array}{cccc}0 & 0 & 0 & 0 \\
0 & 1 & 0 & 0 \\
0 & 0 & 1 & 0 \\
0 & 0 & 0 & 1\end{array}\right)\nonumber
\end{equation}
for $f(t)=0$, and ${\bf K}(t)=0$ otherwise.  Hence, ${\rm ker}~{\bf F}(t)$ is $3$-dimensional for $f(t)=0$, and is $0$-dimensional otherwise.

It follows that for the first consistency condition in equation (\ref{FKcondition}) to be satisfied, $f(t)$ can reach zero at some time $t_0$ only if it remains zero for all later times $t\geq t_0$, i.e., {\it a time-local master equation can exist only if decay to the lower level is irreversible}.  Further, 
noting the form of $\dot{{\bf F}}$ above, it is seen that the second consistency condition in equation (\ref{FKcondition}) is satisfied at $f(t)=0$ if and only if 
\[
\dot{{\bf F}}{\bf K} =\left(\begin{array}{cccc}0 & 0 & 0 & 0 \\
0 & \dot{f}(t) & 0 & 0 \\
0 & 0 & \dot{f}(t) & 0 \\
0& 0 & 0 & 2\dot{f}(t)f(t)\end{array}\right)=0,
\]
that is, if and only if $\dot{f}=0$, and hence, from equation ({\ref{fdotmatrix}), if and only if $\dot{{\bf F}}=0$.  When these conditions are both satisfied, a master equation corresponding to $\phi_t$ exists even when ${\bf F}$ is not invertible. 

In particular, as demonstrated in section \ref{Lsolutionsec}, we may obtain a solution for ${\bf L}$ using the Moore-Penrose inverse of ${\bf F}$, ${\bf L}=\dot{{\bf F}}\tilde{{\bf F}}+{\bf M}$. As noted above, the consistency conditions imply that $\dot{{\bf F}}=0$. Furthermore, the matrix ${\bf F}$ is finite and consequently its singular value decomposition and Moore-Penrose inverse are well-defined. Hence, the general solution for $t\geq t_0$ reduces to ${\bf L}={\bf M}$, where ${\bf M}$ is any matrix satisfying ${\bf M}{\bf F}=0$. We therefore have
\begin{equation}
\label{Meqn}
{\bf L}(t) = \left(\begin{array}{cccc}
	a_1 & b_1 & c_1 &a_1 \\
	a_2 & b_2 & c_2 &a_2 \\
	a_3 & b_3 & c_3 &a_3 \\
	a_4 & b_4 & c_4 &a_4 
	\end{array}\right),
\end{equation}
where $a_i, b_i$ and $c_i$ are arbitrary time-dependent functions, for all $t\geq t_0$. 
Since $\phi_t(\rho)=|1\rangle\langle 1|$ whenever $f(t)=0$, i.e., the atom is in the ground state, it follows that the master equation map corresponding to $\bf L$ in equation (\ref{Meqn}) satisfies $\Lambda(|1\rangle\langle 1|)=0$, irrespective of the choice of functions $a_j(t)$, $b_j(t)$ and $c_j(t)$, with $\Lambda(|2\rangle\langle 2|)$, $\Lambda(|2\rangle\langle 1|)$ and $\Lambda(|1\rangle\langle 2|)$ being arbitrary. Hence we have many possible master equations which generate the same time evolution for $t\geq t_0$.  For earlier times $t<t_0$, where $f(t)\neq 0$, the master equation is of course uniquely specified by equation (\ref{mastereq}) above.

Finally, it is worth pointing out that if the `minimal decoherence' condition (\ref{pf2}) is relaxed, so that the survival probability $p(t)$ and decoherence factor $f(t)$ in equation (\ref{generalqubitevolution}) become independent (subject to the positivity requirement $|f|^2< p$), then we can again calculate the corresponding master equation when the consistency conditions are satisfied. For the case $f(t_0)=0$ but $p(t_0)\neq 0$, the off-diagonal density matrix elements vanish, corresponding to `total decoherence'. In this case, the inverse of $\bf F$ does not exist, and the kernel of $\bf F$ is {\it two}-dimensional. The consistency conditions then imply that $f(t)=0$ for all later times, i.e., that `decoherence is irreversible', and the set of possible master equations for $t\geq t_0$ correspond to those for which $\Lambda(|1\rangle\langle 2|) =0=\Lambda(|2\rangle\langle 1|)$. If at a subsequent time $t_1$ the survival probability becomes equal to zero, this reduces to the case treated above, with the consistency conditions implying that `decay to the lower level is irreversible'.

\subsection{Best possible master equation when no true master equation exists}

As shown in section 5.3, if the consistency conditions are {\it not} satisfied for the above example, then the master equations corresponding to ${\bf L}=\dot{{\bf F}}\tilde{{\bf F}} +{\bf M}$, with ${\bf MF=0}$, minimise the quantity $\| \dot{\phi_t}-\Lambda_t\|$.  Moreover, the particular master equation corresponding to ${\bf L}=\dot{{\bf F}}\tilde{{\bf F}}$ is the `best possible', in the sense of having the smallest possible norm.

In order to calculate this `least squares' solution ${\bf L}$ for the case $f(t)=0$, we have to calculate the singular value decomposition of ${\bf F}$ in equation (\ref{Ft0}), as described in section 5.2.  Only the first singular value is non-zero, being equal to $\sqrt{2}$,  
and the matrices ${\bf U}$ and ${\bf V}$ 
can be chosen as
\begin{equation}
	{\bf U}=\left(\begin{array}{cccc}
	\frac{1}{\sqrt{2}}  & 0 & 0& \frac{1}{\sqrt{2}} \\
	0 & 1 & 0 & 0 \\
	0 & 0 & 1 & 0 \\
	-\frac{1}{\sqrt{2}} & 0 & 0&
	\frac{1}{\sqrt{2}}\end{array}\right),
	~~~{\bf V}= \left(\begin{array}{cccc}1 & 0 & 0 & 0 \\
	0 & 1 & 0 & 0 \\
	0 & 0 & 1 & 0 \\
	0 & 0 & 0 & 1\end{array}\right).
	\nonumber 
\end{equation}
The Moore-Penrose inverse of ${\bf F}$ follows from equation (\ref{moorepenrose}) as 
\begin{equation}
	\tilde{{\bf F}}= {\bf V}\tilde{{\bf D}}{\bf U}^{\rm T} 
	= \left(\begin{array}{cccc}1/2 & 0 & 0 &-1/2 \\
	0 & 0 & 0 & 0 \\
	0 & 0 & 0 & 0 \\
	 0 & 0 & 0 & 0\end{array}\right),
\end{equation}
and hence, using equation (\ref{fdotmatrix}) for $\dot{{\bf F}}$, the `least squares' solution for ${\bf L}$ when $f(t)=0$ is given by
\begin{equation}
\label{Lexamplesolution}
{\bf L} = \dot{{\bf F}}\tilde{{\bf F}}= \left(\begin{array}{cccc}
	0 & 0 & 0 &0 \\
	0 & 0 & 0 & 0 \\
	0 & 0 & 0 & 0 \\
	 \dot{f}(t)f(t) & 0 & 0 & -\dot{f}(t)f(t)\end{array}\right) .
\end{equation}
At all other times, ${\bf L}$ is instead given by equation (\ref{Lmatr}).

If we assume that $\dot{\phi}_t$ is well-defined at all times, then this implies that $\dot{f}(t)$ is finite and therefore that $\dot{f}f=0$ when $f=0$, so that ${\bf L}=0$ when $f=0$. In this case, for ${\bf L}$ as above, one finds that the corresponding minimum value of $\| \dot{\phi_t}-\Lambda_t\|$ in equation (\ref{minimumnorm}) is given by zero whenever $f(t)\neq 0$, and by
\[
\|\dot{\phi_t}-\Lambda_t\| = \|\dot{{\bf F}}\| = \sqrt{2} |\dot{f}(t)|
\]
when $f(t)=0$, where here the last equality follows via equation (\ref{fdotmatrix}).  The distance between the `true' $\dot{\phi}_t$ and our `best guess' $\Lambda_t$ is therefore zero if $\dot{f}=0$ whenever $f=0$.  This corresponds to the second consistency condition in equation (\ref{FKcondition}) being satisfied at all times (for example, the case $f(t)=\cos^2 t$).

However, as noted in section 5.3, even when the second consistency condition is satisfied identically, the corresponding `best possible' master equation can only track the evolution map $\phi_t$ over intervals for which the {\it first} consistency condition in equation (\ref{FKcondition}) is also met, since no first-order equation in time can model the splitting of a trajectory into two or more trajectories.  For example, if  $f(t)=\cos^2 t$, the best possible map $\Lambda_t$ cannot be used to determine $\rho(t_2)$ from $\rho(t_1)$ if the interval $[t_1,t_2)$  includes an odd multiple of $\pi/2$.  It is of interest to note in this regard, using equation (\ref{Lmatr}), that for $t<\pi/2$ the `best possible' master equation map satisfies
\[ \| \Lambda_t \| = \sqrt{10} |\dot{f}(t)/f(t)| = 2\sqrt{5} |\tan t|,\]
and hence diverges in the limit $t \rightarrow \pi/2$, i.e., as $f(t)\rightarrow 0$.

\section{Conclusions}

We have described a general procedure for constructing the linear map corresponding to a given master equation, and vice versa. The master equation is assumed to be local in time, but may otherwise be either Markovian or non-Markovian.
The method involves expressing both the master equation and the linear map in matrix form. Starting from a given master equation, we obtain the Kraus-type representations for the corresponding linear map $\phi_t$. 
The method works whether the linear map $\phi_t$ is completely positive or not. In fact, our construction allows us to {\it determine} whether or not, for a given proposed master equation, the corresponding map is completely positive. It will be completely positive if and only if the eigenvalues of the Choi matrix $\bf S$, corresponding to the linear map, are nonnegative \cite{Choi, caves}. A recent paper gives a different method of obtaining the Kraus representation for the Lindblad equation for Markovian master equations \cite{nakazato}.

Conversely, starting from a linear map $\phi_t$, we ask whether a corresponding master equation, which is local in time, exists. If the (left) inverse of the map $\phi_t$ exists at all times, then the construction of the master equation is straightforward. If the inverse does not exist at a certain time or times, then the situation is more interesting. It is sometimes {\it still} possible to define a master equation which faithfully describes the evolution of the system. 

If the inverse of $\phi_t$ does not exist, then this corresponds to the case when some initial states evolve to the same state at a certain time. Thus, if one is to be able to describe the system with a first-order master equation which is local in time, then these states must follow the same trajectory after this point. This `fidelity constraint' leads to two consistency conditions.
If the map $\phi_t$ satisfies these conditions, then the corresponding master equation can be defined, even when $\phi_t$ is not invertible. The master equation is unique if $\phi_t$ is invertible, otherwise it is not unique.  If the fidelity constraint is not satisfied, meaning that according to $\phi_t$, the trajectories of the states which have `merged' will separate again, then a true master equation, which reproduces the correct time-evolution of the system, and which is first-order and local in time, does not exist. However, even in this case we can define a {\it best possible} master equation $\dot{\rho}=\Lambda_t(\rho)$, in the sense that $\|\dot{\phi}_t- \Lambda_t\|$ and $\|\Lambda_t \|$ are minimal. 
As seen in the example in section 6.3, if, for this `best possible' master equation, one has that $\|\Lambda_t\|$ diverges at some time $t'$, then it  cannot be integrated to approximate $\phi_t$ after this time.

Although master equations have previously been treated by mapping them to matrix equations \cite{BreuerPetroBook}, and the connection between the linear map and its Choi matrix is well known {\cite{Choi, caves}, our intention is here to tie the two concepts together, and show how these tools can be used to go from a master equation to the corresponding linear map and vice versa. This has also led to new insights regarding the existence of time-local master equations even in the case that the corresponding time-evolution is not invertible. We have illustrated the methods by two examples involving two-level quantum systems, discussing the case of invertible and non-invertible $\phi_t$.
In these examples, we met the possibility of Lindblad-type behaviour, loosely speaking corresponding to time-dependent decay rates, but also `anti-Lindblad' behaviour, corresponding to {\it negative} decay rates, or reabsorption and recoherence.

\section*{Acknowledgments}

E. A. would like to acknowledge support from the Royal Society. E. A. and M. H. would like to thank Macquarie University for hospitality.


\begin{thebibliography}{99}
\bibitem{Naka} S. Nakajima, Prog.  Theor.  Phys. \textbf{20} 948 (1958).  	

\bibitem{20} R. Zwanzig, J. Chem.  Phys. \textbf{33} 1338 (1960).	

\bibitem{Chat} S. Chaturvedi and J. Shibata, Z. Physik B \textbf{35}  297 (1979).	

\bibitem{Shib} N. H. F. Shibata, Y. Takahashi, J. Stat.  Phys. \textbf{17} 171 (1977).	

\bibitem{Lindblad} G. Lindblad, Comm.  Math.  Phys. \textbf{48} 119 (1976).	

\bibitem{BreuerPetroBook} H.-P. Breuer and F. Petruccione, {\it The Theory of Open Quantum Systems} (Oxford University Press, Oxford, 2002).

%non-markov error correction
\bibitem{Lidar} R. Alicki, D. A. Lidar and P. Zanardi, Phys. Rev. A \textbf{73} 052311 (2006).
\bibitem{terhal} B. M. Terhal and G. Burkard, Phys. Rev. A {\bf 71}, 012336 (2005).
\bibitem{aharonov} D. Aharonov, A. Kitaev and J. Preskill, Phys. Rev. Lett. {\bf 96}, 050504 (2006).
\bibitem{aliferis} P. Aliferis, D. Gottesman and J. Preskill, Quant. Inf. Comput. {\bf 6} 97-165 (2006).

%trajectory refs
\bibitem{carmichael} H. J. Carmichael, {\it An open systems approach to quantum optics} (Springer-Verlag, Berlin, 1993).
\bibitem{plenio} M. B. Plenio and P. L. Knight, Rev. Mod. Phys. {\bf 70}, 101 - 144 (1998).
\bibitem{kleinekat} U. Kleinekath\"{o}fer, I. Kondov and M. Schreiber, Phys.  Rev. E {\bf 66}, 037701 (2002).
\bibitem{intravaia} F. Intravaia, S. Maniscalco, J. Piilo and A. Messina, Phys. Lett. A {\bf 308} 6 (2003).
\bibitem{Breuer} H.-P. Breuer, B. Kappler and F. Petruccione, Phys.  Rev.  A \textbf{59} 1633 (1999).	
\bibitem{breuer2}  H. P. Breuer, Eur. Phys. J. D {\bf 29}, 105 (2004). 
\bibitem{breuer3} H. P. Breuer, Phys. Rev. A {\bf 69}, 022115 (2004).

\bibitem{Cresser} J. D. Cresser and S. M. Pickles, J. Quant.  and Semiclass. Opt.  \textbf{8} 73-104 (1996).

\bibitem{Budini1} A. A. Budini, Phys. Rev. A \textbf{64} 052110 (2001).	

\bibitem{Budini2} A. A. Budini, Phys. Rev. A \textbf{69} 042107 (2004).	

\bibitem{kraus} K. Kraus, {\it States, effects and operations} (Springer-Verlag, Berlin, 1983). 

\bibitem{Jimphase} J. D. Cresser, unpublished.

\bibitem{Choi} M.-D. Choi, Lin. Alg. Appl. {\bf 10} 285 (1975).

\bibitem{caves} C. M. Caves, J. Supercond. {\bf 12}(6) 707 (1999); also published as quant-ph/9811082.  

\bibitem{diosi} L. Di\'{o}si, N. Gisin, and W. T. Strunz, Phys. Rev. A {\bf 58} 1699 (1998).

\bibitem{hornjohnson} R. A. Horn and C. R. Johnson, {\it Matrix Analysis} (Cambridge University Press, Cambridge, 1985).

\bibitem{barry} B. M. Garraway, Phys. Rev. A {\bf 55} 2290 (1997).

\bibitem{wonderenlendi} A. J. van Wonderen and K. Lendi, J. Stat. Phys. {\bf 100} 633 (2000).

\bibitem{zollerjumps} C. W. Gardiner, A. S. Parkins and P. Zoller, Phys. Rev. A {\bf 46} 4363 (1992). 

\bibitem{dum} R. Dum, P. Zoller and H. Ritsch, Phys. Rev. A {\bf  45} 4879 (1992).

\bibitem{montecarlo1} J. Dalibard,  Y. Castin and K. M\o lmer, Phys. Rev. Lett. {\bf 68} 580 (1992).

\bibitem{montecarlo2} K. M\o lmer, Y. Castin and J. Dalibard, J. Opt. Soc. Am B {\bf 10} 524 (1993).

\bibitem{nakazato} H. Nakazato, Y. Hida, K. Yuasa, B. Militello, A. Napoli, and A. Messina, Phys. Rev. A {\bf 74}, 062113 (2006).

\end{thebibliography}
\end{document}